\documentclass[a4paper, titlepage, 11pt, usenames, dvipsnames, svgnames, table, final]{article}

\usepackage[a4paper, margin=2.5cm]{geometry}

\usepackage{xr}
\usepackage{graphicx}
\usepackage[utf8]{inputenc}

\usepackage{color}

\usepackage{siunitx}
\DeclareSIUnit\kgperL{\kilo\gram\per\liter}
\DeclareSIUnit\waveno{\per\centi\metre}
\DeclareSIUnit\invcm{\per\centi\metre}
\DeclareSIUnit\diffconst{\square\centi\metre\per\second}
\DeclareSIUnit\VpA{\volt\per\angstrom}
\DeclareSIUnit\meVpA{\milli\eV\per\angstrom}

\newcommand{\waveno}{\widetilde{\nu}}

\usepackage[version=3]{mhchem}

\usepackage{caption}
\usepackage{subcaption}

\usepackage{multicol}

\usepackage{verbatim}

\newcommand{\fref}[1]{Fig.~\ref{#1}}
\newcommand{\sref}[1]{Sec.~\ref{#1}}
\newcommand{\eref}[1]{Eq.~\ref{#1}}

\usepackage[hyperref=false,
            url=false,
            isbn=false,
            doi=false,
            eprint=false,
            sorting=none,
            sortcites=true,
            firstinits=true,        
            maxbibnames=99,
            backend=biber]{biblatex}
\AtEveryBibitem
{
    \clearfield{number}
    \clearfield{issue}
    \clearfield{eid}
    \clearfield{note}
}
\DefineBibliographyExtras{english}{}
\DefineBibliographyExtras{english}{}
\bibliography{literature}
\renewbibmacro{in:}{}

\usepackage{authblk}

\makeatletter
\renewcommand*{\@fnsymbol}[1]{\ifcase#1\or*\else\@arabic{\numexpr#1-1\relax}\fi}
\makeatother

\usepackage{bm}
\renewcommand{\vec}[1]{\boldsymbol{\mathbf{#1}}}

\usepackage{pdfpages}

\begin{document}

\title{Probing the Temporal Response of Liquid Water to a THz Pump Pulse Using Machine Learning-Accelerated Non-Equilibrium Molecular Dynamics}

\author[1]{Kit Joll}
\affil[1]{Department of Physics and Astronomy and Thomas Young Centre, University College London, London, WC1E 6BT, United Kingdom}
\author[2,3]{Philipp Schienbein\thanks{email: philipp.schienbein@ruhr-uni-bochum.de}}
\affil[2]{Lehrstuhl für Theoretische Chemie II, Ruhr-Universität Bochum, 44780 Bochum, Germany}
\affil[3]{Research Center Chemical Sciences and Sustainability, Research Alliance Ruhr, 44780 Bochum, Germany}
\date{ }

\maketitle

\begin{abstract}
    Ultrafast, time-resolved spectroscopies enable the direct observation of non-equilibrium processes in condensed-phase systems and have revealed key insights into energy transport, hydrogen-bond dynamics, and vibrational coupling. 
    While ab initio molecular dynamics (AIMD) provides accurate, atomistic resolution of such dynamics, it becomes prohibitively expensive for non-equilibrium processes that require many independent trajectories to capture the stochastic nature of excitation and relaxation. 
    To address this, we implemented a machine learning potential that incorporates time-dependent electric fields in a perturbative fashion, retaining AIMD-level accuracy. 
    Using this approach, we simulate the time-dependent response of liquid water to a 12.3~THz Gaussian pump pulse (1.3~ps width), generating 32~ns of total trajectory data. 
    With access to ab initio-quality electronic structure, we compute absorption coefficients and frequency-dependent refractive indices before, during, and after the pulse. 
    The simulations reproduce key experimental observables, including transient birefringence and relaxation times. 
    We observe energy transfer from the excited librational modes into translational and intramolecular vibrations, accompanied by a transient, nonlinear response of the hydrogen-bond network and a characteristic 0.7~ps timescale associated with librational energy dissipation. 
    These findings demonstrate the method’s ability to capture essential non-equilibrium dynamics  with theoretical time-dependent IR spectroscopy and establish a broadly applicable framework for studying field-driven processes in complex molecular systems.
\end{abstract}

\section{Introduction}

Light-matter interactions are of critical importance in nature, chemical engineering, and fundamental research.
Particularly intriguing is the excitation of a system by an incident light pulse, where the nature of the excitation depends on the light's frequency and polarization.
A widely recognized example is electronic excitation by UV light, which forms the basis for photosynthesis in plants, artificial photosynthesis in solar cells, and photocatalysis at semiconductor interfaces~\cite{Wang-2019-ChemRev, Xu-2019-ChemRev, Franco-2020-ChemSocRev, Yi-2023-NatRevMater, Brinkmann-2024-NatRevMater, Ding-2024-ChemSocRev}.
In contrast, vibrational excitations induced by IR pump pulses~-- which occur on the electronic ground state~-- have received less attention in application-driven contexts, 
despite coupling directly to intramolecular bonds and thus potentially enabling selective bond cleavage~\cite{Morichika-2019-NatCommun}.
This concept has shown promise in gas-phase molecules~\cite{Windhorn-2002-ChemPhysLett, Witte-2003-JCP, Crim-2008-PNAS, Ellerbrock-2022-SciAdv}, but 
achieving such selectivity in the liquid phase remains challenging~\cite{Nuernberger-2007-PCCP, Morichika-2019-NatCommun}.
Further down the energy scale, THz excitations primarily interact with intermolecular modes, such as solvent-solvent interactions in liquid mixtures~\cite{Venables-1998-JCP}, water-water H-bonds in liquid water~\cite{Heyden-2010-PNAS} and solute-solvent interactions in aqueous solutions~\cite{Schienbein-2017-JPCL,Niessen-2019-NatCommun}.
These interactions play a central role in determining solvation properties in the liquid state, 
which in turn influence reaction kinetics~\cite{Slakman-2018-JPhysOrgChem},  
catalytic reaction mechanisms~\cite{Gardner-2024-JACS}, 
and the overall catalytic efficiency~\cite{Riplinger-2014-JACS}.
Solvent molecules can (de)stabilize intermediates and transition states, or coordinate to and block active sites~\cite{Neri-2018-NatCatal, Potts-2021-ChemSocRev}, 
thereby governing key steps of the reaction pathway.
One can therefore speculate that transiently disturbing solvation properties~-- by selectively exciting intermolecular solvent modes using a tailored light pulse~-- might enable active control of chemical reactivity. 
These strategies to selectively control reactions by laser light 
have long been considered among the ``Holy Grails'' in chemistry, 
as they hold the promise of controlling reaction outcomes, enabling novel synthesis techniques, and realizing programmable molecular devices~\cite{Warren-1993-Science,Kohler-1995-AccChemRes}.

When it comes to experimentally monitoring an excitation over time, ultrafast, time-resolved spectroscopies, such as pump-probe and 2D~spectroscopy have been established.
Therein, spectra in the desired frequency regime are collected as a function of time before, during, and after the excitation pulse enabling direct observation of the excited state population as a function of time, and its relaxation back to equilibrium~-- herein, we focus on vibrational spectroscopy only.
Couplings between different vibrational modes can be monitored at pico-second time resolution and it is therefore possible to observe the evolution of non-equilibrium processes as a function of time and across molecular groups~\cite{Nibbering-2005-AnnuRevPhysChem, Hamm-2008-AnnuRevPhysChem}.
This lead to enormously valuable insights on dynamical processes in condensed phase systems~\cite{Zheng-2005-Science, Zhang-2011-NatChem, Rosenfeld-2011-Science}
and specifically 
in liquid water and ice~\cite{Perakis-2016-ChemRev}, for instance energy transport~\cite{Piatkowski-2009-PCCP}, H-bond lifetimes~\cite{Woutersen-2001-ChemPhys, Fecko-2003-Science}, uncovering the collective nature of the H-bond network~\cite{Nicodemus-2011-JPCB, Savolainen-2013-PNAS}, and identifying couplings between different vibrational modes~\cite{Perakis-2012-PCCP}.
Notably, excitations are not limited to vibrational pump pulses, but can be manifold, including temperature-jumps~\cite{Mishra-2018-JPCA}, ultra-fast photoswitches~\cite{Hoberg-2023-ChemSci}, or chemical reactions~\cite{Nibbering-2005-AnnuRevPhysChem}, which can then be observed by time-dependent IR spectroscopy.

Complementing experimental work, theoretical vibrational spectroscopy offers a microscopic view of dynamic processes at the atomistic level.
Molecular dynamics~(MD) simulations
generate atomistic trajectories, enabling a rigorous statistical mechanics analysis that explicitly accounts for thermal fluctuations and anharmonic effects, 
both of which can drastically influence the resulting spectrum.
When studying excitation and corresponding relaxation dynamics, the system is driven into a non-equilibrium state by an external perturbation that injects energy. 
Foundational contributions to this field 
employed classical force fields and hybrid quantum/classical approaches~\cite{Lawrence-2002-JCP, Lawrence-2003-JCP, Rey-2004-ChemRev, Bakker-2010-ChemRev}.
Accordingly, the simulations must be non-equilibrium in nature and require ensemble averaging over many trajectories initiated from different conditions to yield meaningful time-resolved signals~\cite{Evans-2008-NonequilibriumStatMech}.
Over the years, several approaches to modeling excitations have been explored.
One common method involves injecting kinetic energy directly to selected vibrational modes,
either explicitly~\cite{Rey-2009-JPCA, Lesnicki-2018-JPCB}, or via thermostats coupled to specific modes~\cite{Dettori-2017-JCTC, Dettori-2019-JPCL}, 
where the excitation frequency is indirectly determined.
Another strategy to uncover solvent relaxation was to suddenly assign a unit charge to a previously neutral atom in aqueous solution, 
where collective reorganizations and energy transport from that atom could be monitored~\cite{Rey-2015-JPCB}.
A particularly direct method closely reproducing an experimental setup, involves applying an external electric field to mimic the pump pulse. 
The field can either be time-dependent~-- oscillating at specific frequencies with a tailored temporal envelope~\cite{Yagasaki-2011-JCP-134,  Elgabarty-2020-SciAdv}~-- 
or time-localized, i.e., a static field applied for a defined duration~\cite{Yagasaki-2008-JCP, Yagasaki-2009-JCP, Grechko-2018-NatCommun}.
The former approach allows precise excitation-frequency control; the latter is particularly useful for simulating 2D spectra where multiple pump frequencies are involved. 
However, both methods demand many independent simulations to sample the non-equilibrium ensemble adequately. 
For instance, calculating a 2D IR spectrum of liquid water required over $10^6$ trajectories of 250~fs each (totaling 250~ns) to achieve convergence~\cite{Grechko-2018-NatCommun}.

It is particularly desirable to run these simulations with \emph{ab initio} MD~(AIMD) techniques, 
in which the electronic structure is computed on-the-fly at every time step. 
This enables the inclusion of electronic polarization and charge transfer effects, providing highly accurate and often predictive vibrational spectra that can quantitatively match experimental results~\cite{Mauelshagen-2025-SciAdv}.
Most importantly, AIMD simulations allows chemical reactions to occur naturally within the simulation, making it particularly relevant when investigating reaction steering as a consequence of vibrational excitation.
Due to the computational cost, most non-equilibrium pump-probe simulations have relied on classical force field MD (FFMD) simulations~\cite{Imoto-2015-JPCB, Ojha-2019-CommChem, Dettori-2017-JCTC, Mishra-2018-JPCA, Dettori-2019-JPCL, Elgabarty-2020-SciAdv, Ojha-2021-SciRep}, including more advanced polarizable force field models, such as AMOEBA~\cite{Novelli-2020-JPCB}.
While, AIMD has been successfully employed in selected cases~-- particularly when excitation pulses are short and fast processes are probed~\cite{Nagata-2015-PRX, Mishra-2018-JPCA, Elgabarty-2020-SciAdv}~--
it remains resource-intensive and typically limited to GGA-level functionals.
This becomes particularly limiting when more accurate levels of electronic structure are needed.
As a counterexample, experimental studies have employed Gaussian pump pulses with widths on the order of \num{3}~ps~\cite{Novelli-2020-JPCB, Novelli-2022-PCCP}.
In such cases, simulations must span at least 5~ps to cover the full pulse and include the relaxation back to equilibrium~-- often requiring even longer trajectories. 
To achieve statistical convergence, multiple such simulations are needed, and the total trajectory time can accumulate to several nanoseconds. 
Under these conditions, full AIMD becomes prohibitively expensive, motivating the development of accelerated approaches that retain ab initio accuracy while enabling access to longer timescales.

With the advent of machine learning, AIMD simulations can be accelerated to retain ab initio accuracy at only a fraction of the computational cost~\cite{Behler-2007-PRL, Bartok-2010-PRL, Schuett-2017-NatCommun, Behler-2021-ChemRev, Batzner-2022-NatCommun}.
In these approaches, energies and forces obtained from explicit electronic structure calculations are used to train models that can then reproduce them efficiently during machine learning molecular dynamics (MLMD) simulations.
However, many ML potentials focus solely on reproducing forces and energies, and do not include electronic properties such as dipole moments or polarizabilities~-- quantities that are essential for computing vibrational spectra. 
In recent years, several extended models have been developed to address this limitation by 
incorporating these additional properties directly into the learning targets or by training them separately~\cite{Gastegger-2017-ChemSci, Sifain-2018-JPCL, Wilkins-2019-PNAS, Unke-2021-NatCommun, Schuett-2021-ProcMachineLearningRes, Gao-2022-NatCommun, Cools-2022-JCTC, Beckmann-2022-JCTC, Schienbein-2023-JCTC}, 
enabling the calculation of vibrational spectra within MLMD~\cite{Han-2022-JPCA}.
To this end, we recently introduced the atomic polar tensor~\cite{Person-1974-JCP}~-- also known as Born effective charge tensor~\cite{Gonze-1992-PRL}~-- as a machine learning target that provides access to accurate IR spectra~\cite{Schienbein-2023-JCTC}.
A key advantage of the atomic polar tensor is that it is a rigorous physical observable, well-defined for each atom, and does not require artificial charge partitioning.
Building on this, we further demonstrated that the atomic polar tensor can be used to incorporate the response to an external electric field directly into MLMD simulations~\cite{Joll-2024-NatCommun}. 
We implemented this field response as a perturbation, where the total force is split into two contributions:
(1) the unperturbed force from a standard machine learning interatomic potential, and
(2) the perturbative force arising from the electric field, as predicted by the atomic polar tensor.
These contributions are treated independently, with their corresponding models kept separate, and the electric field dependence only enters during the MLMD simulation.
We thus termed this approach the perturbed neural network potential (PNNP).
We demonstrated that this framework can faithfully describe liquid water under electric fields strengths up to about \SI{0.2}{\VpA}, which is substantial given that rare spontaneous dissociation events occur at higher fields~\cite{Joll-2024-NatCommun}.
The PNNP is thus particularly well suited for studying time-dependent pump-probe vibrational spectroscopy, as it enables the explicit modeling of the excitation pump pulse via MLMD. 
Moreover, the required electronic
observables
to compute the time-dependent IR spectrum~-- namely, the atomic polar tensors~-- are already available within the simulation, as they are used to account for the perturbative force exerted by the external field.

In this work, we calculate the time-dependent response of liquid water to an incident THz pump pulse, closely matching a recent experimental study~\cite{Novelli-2020-JPCB, Novelli-2022-PCCP}.
We perform PNNP simulations to model the pump pulse and monitor the relaxation as a function of time at \emph{ab initio} accuracy.

\section{Methods}

To model the electric field pulse we now employ the recently introduced PNNP approach~\cite{Joll-2024-NatCommun}, which builds on the idea of training atomic polar tensors~\cite{Schienbein-2023-JCTC}.
We employ a committee~\cite{Schran-2020-JCP} of second-generation high-dimensional neural network potentials~(HD-NNP)~\cite{Behler-2007-PRL, Behler-2021-ChemRev} for the interatomic potential, while an atomic polar tensor neural network~(APTNN) is used to obtain the required atomic polar tensors~\cite{Schienbein-2023-JCTC}, based on \texttt{e3nn}~\cite{Geiger-2022-arxiv, e3nn} and \texttt{pytorch}~\cite{pytorch}.
The employed networks are taken unchanged from our previous publications~\cite{Schienbein-2023-JCTC, Joll-2024-NatCommun}, so no additional training was necessary in this work.
Both models were trained at the DFT level, using the RPBE functional~\cite{Hammer-1999-PRB} supplemented by D3 dispersion corrections~\cite{Grimme-2010-JCP}.
This electronic structure method has shown to reliably model fluid water, reproducing a wide range of experimental properties with near-quantitative accuracy, including vibrational spectra, 
across the phase diagram~-- 
from ambient conditions to high pressures and high temperatures in the supercritical state~\cite{ForsterTonigold-JCP-2014, Imoto-2015-PCCP, Morawietz-2016-PNAS, Schienbein-2020-PCCP, Schienbein-2020-ANIE, Noetzel-2024-ANIE, Mauelshagen-2025-SciAdv}~-- 
as well as liquid water at metal interfaces~\cite{Gross-2022-ChemRev}.
Note that the previously trained models for the intermolecular potential and the atomic polar tensors are publicly available online~\cite{Joll-2024-NatCommun}.
Instead of using a static electric field, we now proceed and apply a time-dependent electric field pulse mathematically formulated as:
\begin{equation}
    E(t) = 
    E_0 
    \cos\left(\omega_{\text{e}} t + \varphi\right)
    \exp\left(-\frac{(t-t_0)^2}{2\sigma^2}\right)
    \, ,
    \label{eq:pump}
\end{equation}
which mimics
the electric field part of a monochromatic light pulse at frequency $\omega_{\text{e}}$, phase $\varphi$, and maximum amplitude $E_0$, 
and that
is confined in time by a Gaussian envelope, where $t_0$ is the center of the Gaussian in time and $\sigma$ is the associated width.
We previously~\cite{Joll-2024-NatCommun} implemented the PNNP framework for static electric fields in \texttt{CP2k}~\cite{Kuehne-2020-JCP} and have extended it here 
to support time-varying electric fields.
In the following an electric field amplitude of $E_0=\SI{0.027}{\VpA}$, a frequency of $\omega_{\text{e}} = 2\pi \times 12.3$~THz and a Gaussian width of $\sigma=\num{1.3}$~ps is employed, comparable to  recent experiments~\cite{Novelli-2020-JPCB, Novelli-2022-PCCP}, where the same frequency and similar pulse widths and electric field strengths were used.
We begin the data generation by running a single NVT simulation without an applied field for 1~ns after equilibration.
From the  NVT simulation we sample 800 configurations, separated by 1.25~ps each, which are used as initial conditions for subsequent simulations including the pump pulse.
For each of these initial conditions, we run a 20~ps simulation in the NVE ensemble with the pump pulse applied, centered at $t_0=$~5~ps, ensuring the width of the  pump pulse is fully sampled and giving the system sufficient time to relax back to equilibrium; 
the temporal profile of the pump is shown in \fref{fig:pulse_and_temp} (a).
To account for the effect of the phase $\varphi$, 800 simulations were generated using $\varphi = 0$ and another 800 simulations were generated using $\varphi = \pi/2$ using the same set of initial conditions, resulting in a total simulation time of 32~ns, using a time step of 0.5~fs.
By averaging the trajectories from both phases, the phase dependence of the external field is eliminated~\cite{Imoto-2015-JPCB}. 
All simulations were performed using the \texttt{CP2k} program package~\cite{Kuehne-2020-JCP}.

The time dependent absorption coefficient 
\begin{equation}
    \alpha(\tau_2, \omega) n(\tau_2, \omega)
    = 
    \frac{\beta\pi}{3 c V \epsilon_0} 
    \int_{-\infty}^{\infty} e^{-i\omega t}
    \left<\dot{\vec{M}}(\tau_2)\dot{\vec{M}}(\tau_2 + t) \right>
    dt
    \label{eq:absorption}
\end{equation}
is computed
from time correlation function of the total dipole moment time derivative $\dot{\vec{M}}(t)$, 
where $\omega$ is the angular (``probe'') frequency, 
$n(\tau_2, \omega)$ the frequency dependent refractive index at pump-probe delay $\tau_2$, 
$c$ the speed of light in vacuum,
$V$ the volume, 
$\epsilon_0$ the vacuum permittivity, 
$\beta = 1/k_\text{B}T$, 
$k_\text{B}$ the Boltzmann constant, 
and 
$T$ is the temperature. 
The time-dependence of the absorption coefficient enters through the two-point time-correlation function,
which expresses the correlation of $\dot{\vec{M}}(t=\tau_2)$ forward in time. 
$\tau_2$ is the pump-probe delay, thus the time relative to the maximum of the pump pulse $t_0$, see \eref{eq:pump}.
Positive values of $\tau_2$ correspond to times \emph{after} the pulse maximum, while negative times represent times \emph{before} it~-- as commonly done in the literature.
A key statistical challenge in this non-equilibrium setting is that the dependence on $\tau_2$ significantly reduces the number of initial configurations available for averaging, compared to an equilibrium simulation. 
To address this, we run a large number of independent trajectories (1600 in total, see above) and further average over a short time window of 0.4~ps centered around each $\tau_2$, as described in detail in the Supporting Information.
Due to the explicit time-dependence through the two-point correlation function, the refractive index $n(\tau_2, \omega)$ also becomes time-dependent. 
We compute $n(\tau_2, \omega)$ using the Kramers-Kronig relations, as described elsewhere~\cite{Iftimie-JCP-2005, Sun-2014-JACS}, and separate it from the absorption coefficient $\alpha(\tau_2, \omega)$.

An advantage of having atomic polar tensors available along the trajectory is that the spectrum can rigorously be decomposed into atomistic contributions by inserting the definition of the atomic polar tensor into \eref{eq:absorption}
\begin{equation}
    \left<\dot{\vec{M}}(\tau_2)\dot{\vec{M}}(\tau_2 + t) \right>
    =
    \left< \sum_i \vec{P}_i(\tau_2) \vec{v}_i(\tau_2) \cdot \sum_j \vec{P}_j(\tau_2 + t) \vec{v}_j(\tau_2 + t) \right>
    \equiv
    \sum_{ij}
    C_{ij}(\tau_2, t)
    \label{eq:apt-decomp}
\end{equation}
such that the time correlation function now depends on the atomic polar tensors $\vec{P}_i(t)$ and velocities $\vec{v}_i(t)$ of the atoms $i$ and $j$ at time $t$, where both sums run over all atoms~\cite{Schienbein-2023-JCTC}.
Since both, the correlation function and the Fourier transform in \eref{eq:absorption} are linear, the sums can be taken out and the absorption coefficient 
\begin{equation}
    \alpha(\tau_2, \omega) 
    = \sum_{ij} \alpha_{ij}(\tau_2, \omega) 
    = 
    \sum_{ij}
    \frac{\beta\pi}{3 c V \epsilon_0 n(\tau_2,\omega)} 
    \int_{-\infty}^{\infty} e^{-i\omega t}
    C_{ij}(\tau_2, t) \,
    dt
    \label{eq:apt-decomp-alpha}
\end{equation}
can be readily expressed as a sum of atomic contributions.
Notably, this is reminiscent of the ``cross correlation analysis'', where the absorption coefficient was expressed in terms of molecular dipole moments~\cite{Schienbein-2017-JPCL}, but goes significantly beyond, since the definition of ``molecules'' is not required any more and therefore bypasses complications when the absorption coefficient of side chains at a single molecule are calculated~\cite{Esser-2018-JPCB}.
Furthermore, such an atomic decomposition has been attempted previously in the literature, however, due to the enormous computational demand to explicitly calculate atomic polar tensors without a ML model as proxy, 
several approximations, ranging in severity, were necessary,
such as the instantaneous normal mode analysis~\cite{Buchner-1992-JCP, Kindt-1997-JCP, Imoto-2019-JCP}, parametrizing the atomic polar tensor~\cite{Khatib-2017-JPCL}, or computing the atomic polar tensor not at every time step~\cite{Gaigeot-2007-MolPhys}.

\section{Results}

Previously, we demonstrated the accuracy of the PNNP model for simulating liquid water under static, time-independent electric fields up to \SI{0.2}{\VpA}, achieving a total force RMSE of approximately \SI{90}{\meVpA}~\cite{Joll-2024-NatCommun}.
We further showed excellent agreement between the PNNP model and explicit AIMD simulations in reproducing the dielectric constant, IR spectra, and reorientational relaxation times across the investigated electric field strengths.
Having established this accuracy for static fields, we now aim to demonstrate that the same PNNP model can accurately describe time-varying electric fields as well, i.e.\ excitation pulses as described by \eref{eq:pump}.
As a commonly used metric in the ML community, we begin by evaluating the root-mean-square error (RMSE) on an independent test set. 
This test set consists of 750 configurations randomly selected from the generated trajectories. 
The electronic structure for each test configuration was computed using \texttt{CP2k}~\cite{Kuehne-2020-JCP} and the \texttt{Quickstep} module~\cite{VandeVondele-2005-ComputPhysCommun},
following the exact same protocol used to train the ML models (c-NNP and APTNN) as described in our previous works~\cite{Schienbein-2023-JCTC, Joll-2024-NatCommun}.
From this test set, we obtain a total force RMSE of \SI{90.9}{\meVpA} consistent with our previous work~\cite{Joll-2024-NatCommun} and comparable to other studies on a wide variety of systems~\cite{Schran-2020-JCP, Schran-2021-PNAS}, indicating that liquid water under the influence of a now time-dependent external electric field is faithfully described by the PNNP model.
\fref{fig_si:absorption_coefficient} and \fref{fig_si:refractive_index} in the Supporting Information present the full IR absorption coefficient $\alpha(\tau_2, \omega)$ and the frequency-dependent refractive index $n(\tau_2, \omega)$, respectively, at pump-probe delay times $\tau_2 = \num{-5}$, \num{0}, and \num{5}~ps, corresponding to times before, during, and after the pump.
For the equilibrium case before the pump pulse (i.e., at $\tau_2=\num{-5}$~ps), we additionally show a comparison with available experimental data at ambient conditions~\cite{Bertie-1996-ApplSpectrosc}.
The very good agreement for both, $\alpha(\tau_2, \omega)$ and $n(\tau_2, \omega)$ highlights the accuracy of the present method in reproducing the optical properties of liquid water across the full frequency range from 10 to \SI{4500}{\invcm}~-- including the excitation frequency of 12.3~THz ($\approx \SI{410}{\invcm}$).

\newcommand{\da}{\Delta \alpha_{\omega_\text{e}}(\tau_2)}
\newcommand{\dan}{\Delta \left[\alpha_{\omega_\text{e}}(\tau_2) n_{\omega_\text{e}}(\tau_2)\right]}

\begin{figure}
    \centering
    \includegraphics[width=0.45\textwidth]{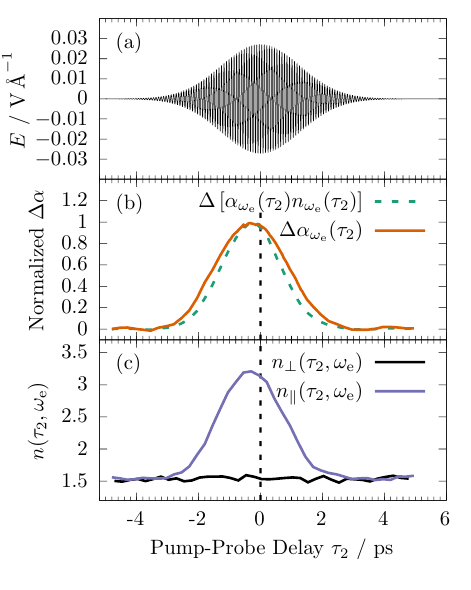}
    \caption{
        (a) Illustration of the applied time-dependent electric field mimicking the pump pulse, as defined in \eref{eq:pump}, shown as a function of the pump-probe delay $\tau_2$.
        (b) Time-dependent differences $\da$ (\eref{eq:diff_a}, orange solid line) and $\dan$ (\eref{eq:diff_an}, green dashed line), representing the changes in absorption at the excitation frequency $\omega_\text{e}$ (\num{12.3}~THz) relative to equilibrium conditions.
        (c) Time-dependent refractive index at $\omega_\text{e}$, plotted parallel ($n_\parallel$, blue solid line) and perpendicular ($n_\perp$, black solid line) to the polarization direction of the applied electric field.
        The vertical black dashed line in (a) and (b) indicates $\tau_2 = 0$~ps.
    }
    \label{fig:pulse_and_temp}
\end{figure}

In \fref{fig:pulse_and_temp} we present the employed pump pulse in panel (a) and the resulting temporal evolution of the absorption coefficient at the excitation frequency $\omega_\text{e}$ (\num{12.3}~THz) in panel (b).
Specifically, we calculate the differences
\begin{equation}
    \da = \alpha(\tau_2, \omega_{\text{e}}) - \alpha_{\text{eq}}(\omega_{\text{e}})
    \label{eq:diff_a}
\end{equation}
and
\begin{equation}
    \dan = \alpha(\tau_2, \omega_\text{e}) n(\tau_2, \omega_\text{e}) - \alpha_\text{eq}(\omega_\text{e}) n_\text{eq}(\omega_\text{e})
    \label{eq:diff_an}
\end{equation}
with respect to their respective equilibrium values 
$\alpha_{\text{eq}}(\omega_{\text{e}})$
and
$\alpha_\text{eq}(\omega_\text{e}) n_\text{eq}(\omega_\text{e})$, 
respectively.
We then normalize them by their corresponding maxima in time which is presented in \fref{fig:pulse_and_temp}.
We observe the typical increase in intensity at negative delay times, due to continuous excitation by the pump pulse, followed by a relaxation back to equilibrium at positive delay times. 
Remarkably, the Gaussian-like shape of $\da$ is visibly broader than that of $\dan$, implying that the frequency dependent refractive index $n(\tau_2, \omega_\text{e})$ has a significant effect and should not be neglected.
This is supported by panel (c), where $n(\tau_2, \omega_\text{e})$ is plotted both perpendicular ($n_\perp$) and parallel ($n_\parallel$) to the polarization of the applied pump pulse.
We observe that $n_\parallel$ increases significantly, approximately doubling at its peak compared to the equilibrium value. 
Interestingly, $n_\perp$ remains essentially unchanged, indicating that the THz pulse primarily affects the parallel component.
The birefringence~-- defined as the difference between the parallel and perpendicular refractive indices~-- reaches an unrealistically large value (see discussion below) of approximately 
$\Delta n = n_\parallel - n_\perp = \num{1.64}$ at the peak of the pump pulse, 
which is also much larger than the reported value in the corresponding experiments~\cite{Novelli-2020-JPCB, Novelli-2022-PCCP}.
This implies that liquid water becomes strongly optically anisotropic, a phenomenon that has indeed been observed experimentally at the same excitation frequency 
$\omega_\text{e}$ used here~\cite{Novelli-2020-JPCB, Novelli-2022-PCCP}.
This effect further forms the basis of optical Kerr-effect spectroscopy, which measures the change in birefringence as a function of pump-probe delay time~\cite{Hoffmann-2009-ApplPhysLett, Elgabarty-2020-SciAdv}.
Alongside the described birefringence, we also fully reproduce the experimentally observed dichroism~\cite{Novelli-2020-JPCB, Novelli-2022-PCCP}, characterized by increased absorption for probe light polarized parallel to the pump field compared to perpendicular polarization. 
This anisotropic absorption behavior is clearly reflected in the spectra calculated here and shown in \fref{fig_si:absorption_coefficient}b in the Supporting Information.

To interpret the spectral response, we view the monitored intensity as a measure for the population of excited states. 
This allows us to apply a simple kinetic model, in which the excitation pulse (\eref{eq:pump}) promotes vibrational excitation, while the excited population decays via first-order kinetics. 
For the excitation we assume zeroth-order kinetics as there are plenty of ground states available and their depletion thus does not limit excitation.
Similar ideas to use a kinetic model to express the pump-probe signal have been employed previously in mid-IR pump-pulse spectroscopy as well~\cite{Woutersen-1997-JCP, Lock-2001-JPCA, Lindner-2006-ChemPhysLett, Seehusen-2009-PCCP, Yagasaki-2011-JCP-134}.
The resulting differential equation,
\begin{equation}
    \frac{dN(t)}{dt} = k_e I(t) - k_r N(t) \, ,
\end{equation}
can be integrated and fitted to the computed temporal response, 
where $N(t)$ is the population of excited states, 
$k_e$ and $k_r$ are the excitation and relaxation rate constants, respectively,
and 
$I(t) = E(t)^2$ is the intensity of the pump pulse, derived from the applied electric field, see \eref{eq:pump}.
Remarkably, we find $k_r^{-1} = \SI{0.79}{\pico\second}$ for the relaxation of $\da$ and $k_r^{-1} = \SI{0.25}{\pico\second}$ for $\dan$, reflecting the visibly slower relaxation in \fref{fig:pulse_and_temp}~(b).
Typically the transmission is reported experimentally as a function of the waiting time $\tau_2$, which is linked to $\alpha(\omega)$ according to the Beer-Lambert law, not the product $\alpha(\omega)n(\omega)$.
Therefore, only the determined relaxation rate from $\da$ is directly comparable to experimental data, which we will do in the discussion.
Finally, we also observe that the peak of the response occurs  at a waiting time of about $\tau_2 \approx \SI{-0.2}{\pico\second}$, indicating that the maximum signal precedes the center of the pump pulse.

\begin{figure}
    \centering
    \includegraphics[width=0.45\textwidth]{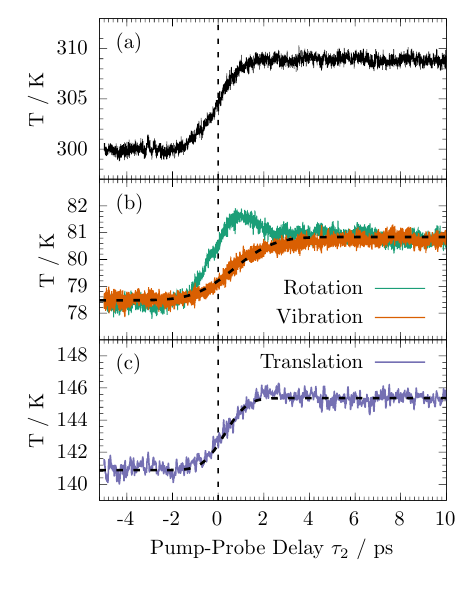}
    \caption{
        (a) Simulation temperature as a function of the pump-probe delay $\tau_2$.
        (b) Rotational (green solid line) and vibrational (orange solid line) contributions to the simulation temperature, based on the decomposition of the atomic velocities (see \eref{eq:velocity-decomp}).
        (c) Translational (blue solid line) contribution to the simulation temperature, computed analogously.
        The vertical black dashed line indicates $\tau_2 = 0$~ps.
        Note that the equipartition theorem only holds for the total velocity of each atom $\vec{v}_i(t)$, i.e.\ sum of translations, rotations, and vibrations, and not for the separated contributions.
    }
    \label{fig:temperatures}
\end{figure}

In panel (a) of \fref{fig:temperatures}, we show the average temperature increase as a function of pump--probe delay time in response to the pump pulse.
We observe a temperature rise, averaged over all trajectories, of about 9~K. 
While this leads to a measurable difference in the equilibrium, time-independent THz spectrum~\cite{Zelsmann-1995-JMolStruct}, the change remains quantitatively minor and within the statistical error bars
of the spectra computed here, despite a total sampling time of 
32~ns,
see \fref{fig_si:absorption_coefficient} in the Supporting Information.
To understand the microscopic response, we separate the velocities into contributions to molecular translations, rotations, and vibrations,
\begin{equation}
    \vec{v}_i(t) = \vec{v}_{i,\text{t}}(t) + \vec{v}_{i, \text{r}}(t) + \vec{v}_{i, \text{v}}(t) \, ,
    \label{eq:velocity-decomp}
\end{equation}
where $\vec{v}_i(t)$ is the velocity of atom $i$ at time $t$ and 
$\vec{v}_{i,\text{t}}(t)$, $\vec{v}_{i, \text{r}}(t)$, and $\vec{v}_{i, \text{v}}(t)$
are the decomposed contributions to the translation, rotation, and internal vibrations of its corresponding water molecule.
Indeed, such a decomposition has been frequently used in the literature for liquid water~\cite{Buchner-1992-JCP, Cho-1994-JCP, Yagasaki-2011-JCP-134, Imoto-2019-JCP, Elgabarty-2020-SciAdv}.
We refer the interested reader to these references for the corresponding mathematical foundation~-- for example, see Section~III~B in Ref.~\cite{Imoto-2019-JCP}.
We present the  average temperatures corresponding to the rotations and vibrations in panel (b) and to the translations in panel (c) of \fref{fig:temperatures}.
Note that the equipartition theorem holds for the total velocity of each atom $\vec{v}_i(t)$, i.e.\ sum of the three contributions, and not for the separated contributions.
We observe that the kinetic energy contained in the water rotations shows the fastest increase.
This is expected since the pump pulse directly excites the librational band, i.e.\ the hindered rotations of the water molecules. 
The rotational kinetic energy experiences a maximum around $\tau_2 = $~1~ps, before it slowly decays to its equilibrium value, implying that the excitation rate is larger than the energy transfer rate from the rotations to the translations and vibrations. 
In case of the translational energy, we observe an error-function-like shape beginning at about \num{-1}~ps until 2~ps, where it reaches its equilibrium value.
Fitting an error-function to the data, we find a time constant of \num{1.08}~ps.
Doing the same for the vibrational energy in panel (b), we find a time constant of \num{1.96}~ps, indicating that the energy transfer from the rotations into the translational modes is about twice as fast as the transfer to the vibrational modes.
Since the excitation primarily targets the rotational modes, we infer that the increase in translation and vibrational kinetic energy can only be caused indirectly by the decay of the rotational modes.
The associated latter rate constant can be estimated by 
$\tau_r = 1/\left(\tau_t^{-1} + \tau_v^{-1}\right) = \num{0.70}$~ps, 
which is remarkably close to the time constant extracted from the relaxation of the absorption coefficient in \fref{fig:pulse_and_temp} at the excitation frequency.

\begin{figure}
    \centering
    \includegraphics[width=0.45\textwidth]{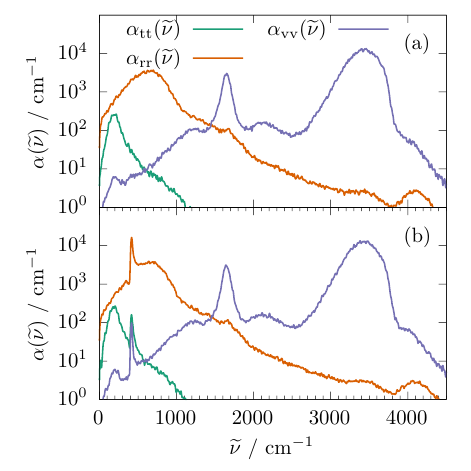}
    \caption{
        (a) The IR spectrum of liquid water at 300~K under equilibrium conditions, and (b) at a pump-probe delay time of 0~ps (i.e., at the peak of the pump pulse), 
        decomposed into translational (green), rotational (orange), and vibrational (blue) contributions according to \eref{eq:apt-decomp-alpha}.
        The full (not decomposed) IR absorption coefficients at pump-probe delay times of \num{-5}, \num{0}, and \num{5}~ps are shown in \fref{fig_si:absorption_coefficient} in the Supporting Information.
    }
    \label{fig:spectra-trv}
\end{figure}

Having atomic polar tensors available along the trajectory, we can now proceed and calculate the absorption coefficient individually for translational, rotational, and vibrational motion.
Inserting \eref{eq:velocity-decomp} into \eref{eq:apt-decomp}, 
as outlined in \sref{sec_si:atomic-decomp} in the Supporting Information
leads to
\begin{equation}
    \alpha(\tau_2, \omega) = \sum_{\zeta\xi} \alpha_{\zeta\xi}(\tau_2,\omega)
    \, ,
    \label{eq:absorption-trv}
\end{equation}
where $\zeta, \xi \in \{\text{t, r, v}\}$,
indicating if the absorption coefficient has been calculated using the translational, rotational, or vibrational velocity contribution, respectively.
Note that we average over all atoms, such that the corresponding sums over atoms $i$ and $j$ (see \eref{eq:apt-decomp-alpha}) vanish.
We present the decomposed spectra in equilibrium at \num{300}~K in \fref{fig:spectra-trv}~(a).
We observe that the well-known absorption bands of liquid water in the THz and mid-IR frequency regime are indeed well separated in the three terms:
The translational spectrum $\alpha_\text{tt}(\omega)$ contains the H-bond stretch vibration centered around \SI{200}{\invcm}, 
the rotational spectrum $\alpha_\text{rr}(\omega)$ contains the librational band around \SI{600}{\invcm}, 
and the vibrational spectrum $\alpha_\text{vv}(\omega)$ contains the intramolecular HOH bending and the intramolecular OH stretch vibrations at around \num{1600} and \SI{3500}{\invcm}, respectively.
This assignment aligns well with the common assignment of absorption bands in liquid water~\cite{Bertie-1996-ApplSpectrosc}.
Remarkably, only by taking the atomic polar tensor explicitly into account, the H-bond stretch vibration around \SI{200}{\invcm} is visible.
In stark contrast, if the VDOS of the translational velocity is calculated, i.e.\ neglecting the atomic polar tensor and thus considering the vibrational density of states only, this band remains invisible~\cite{Heyden-2010-PNAS}.
In panel (b) of \fref{fig:spectra-trv}, we show the same contributions, but at a waiting time of $\tau_2 = \SI{0}{\pico\second}$, i.e.\ at the maximum of the pump pulse.
We observe an additional peak at around \num{12.3}~THz ($\approx \SI{410}{\invcm}$) corresponding to the frequency of the pump pulse, $\omega_\text{e}$.
The highest intensity is present in the rotational contribution of the water molecules.
This is sensible as the librational band is primarily excited by the pump pulse.
Remarkably, we observe a similar peak in the translational and vibrational contributions as well, however, the absolute intensity  is still two orders of magnitude smaller~-- note the logarithmic scale of the plot.
These peaks suggest non-zero cross couplings between translations, rotations, and vibrations.
Such cross couplings imply that the excitation pulse, although clearly targeted at the librational band and thus rotations, acts on the translations and vibrations in the system as well, due to intimate electronic couplings between the different kinds of motion.
This can further potentially explain the large heating rates of translational and vibrational kinetic energies, since these electronic couplings between rotations and translations and vibrations impose a fast heat dissipation pathway.

\begin{figure}
    \centering
    \includegraphics[width=0.45\textwidth]{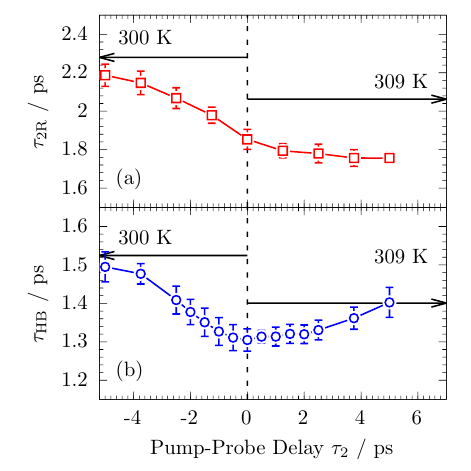}
    \caption{
        (a) Reorientational relaxation time $\tau_\text{2R}$ (red rectangles) and (b) continuous H-bond lifetimes (blue circles) as a function of the pump-probe delay $\tau_2$.
        Error bars represent the standard deviation obtained from block averaging.
        The solid lines connecting the data points are included as visual guides only.
        Arrows pointing to the left and right indicate the respective equilibrium values at 300 and 309~K.
    }
    \label{fig:dynamics}
\end{figure}

Finally, we assess how the pump pulse affects the dynamics of liquid water. 
Of particular relevance is the hydrogen-bond (H-bond) lifetime~\cite{Rapaport-1983-MolPhys, Luzar-1996-Nature}, which quantifies the dynamic stability of the extended H-bond network. 
Closely related is the reorientational relaxation time $\tau_{2R}$~\cite{Matubayasi-2001-JCP}, representing the average time before a water molecule loses memory of its initial orientation.
Although these two quantities capture different physical processes, they are inherently linked: a water molecule must break an H-bond before it can fully reorient, and conversely, if a water molecule fully reorients, an H-bond is likely broken.
We have previously used these two metrics in concert to demonstrate the absence of H-bonding in supercritical water from a dynamical perspective~\cite{Schienbein-2020-ANIE, Mauelshagen-2025-SciAdv}.
Notably, the reorientational relaxation time is a physical observable that can be measured directly using NMR spectroscopy~\cite{Matubayasi-2001-JCP}.
The situation is more nuanced for the H-bond lifetime, as its precise value depends on the specific geometric or energetic definition employed. 
In this work, we use a joint angle--distance criterion established previously~\cite{Imoto-2015-PCCP}, which has been shown to yield reliable estimates of H-bond-related properties in fluid water and aqueous solutions~\cite{Schienbein-2017-JPCL, Schienbein-2020-PCCP, Schienbein-2020-ANIE, Mauelshagen-2025-SciAdv}.

In \fref{fig:dynamics} we present the time-dependent evolution of the reorientational relaxation time $\tau_\text{2R}$ (panel a) and the H-bond lifetime $\tau_\text{HB}$ (panel b).
Both quantities are computed up to a delay time of \num{5}~ps only. 
This limitation arises because they are calculated from forward time-correlation functions 
(see \sref{sec_si:tcf})
which must decay to zero within the available simulation window. 
Thus, sufficient simulation time after each $\tau_2$ value is necessary, and longer simulations would be required to obtain reliable data at later delay times.
The equilibrium values at 300 and 309~K are shown for reference. 
The 300~K data were obtained from explicit equilibrium simulations using the same c-NNP model employed in the non-equilibrium simulations, while the 309~K value was derived by interpolating established literature data between 300 and 350~K~\cite{Schienbein-2020-PCCP, Schienbein-2020-ANIE}.
From these equilibrium values, we observe that both, $\tau_\text{2R}$ and $\tau_\text{HB}$ are expected to decrease with increasing temperature~-- a behavior that only reflects a change of the thermodynamic state point and is not specific to the application of a THz pulse.
Inspecting panel (a), we find that $\tau_\text{2R}$ decreases monotonically up to a pump-probe delay of \num{5}~ps, with an inflection point around 0~ps, coinciding with the peak of the pump intensity.
Remarkably, even at \num{5}~ps, the equilibrium value expected at 309~K has not yet been reached.
Although $\tau_\text{2R}$ and $\tau_\text{HB}$ related, the temporal behavior of $\tau_\text{HB}$ is qualitatively different, as shown in panel (b). 
Initially, it decreases from 1.49~ps until $\tau_2 = \num{0}$~ps, reaching a minimum at \num{1.30}~ps, and then increases again, approaching its equilibrium value of \num{1.40}~ps around $\tau_2=\num{5}$~ps. 
Since $\tau_\text{2R}$ has not yet reached its equilibrium value at \num{5}~ps, it is reasonable to suspect that it continues to increase at later delay times.
Overall, the markedly non-linear response of the H-bond dynamics suggests that the extended hydrogen-bond network is transiently but significantly perturbed by the applied THz pump pulse. 
Since this response differs from the behavior observed when simply heating the liquid from 300 to 309 K, it appears to be specific to the characteristics of the THz excitation~-- both in intensity and temporal profile.

\section{Discussion}

We estimate a relaxation time of $k_r^{-1} = 0.79$~ps of the absorption coefficient $\Delta\alpha_{\omega_\text{e}}(\tau_2)$ (\fref{fig:pulse_and_temp}b) at 12.3~THz (\SI{410}{\invcm}) coinciding with the librational band of liquid water.
This time scale is a recurring feature of liquid water, consistently observed in experimental pump-probe experiments, covering mid-IR and THz excitations~\cite{Lock-2001-JPCA, Cowan-2005-Nature, Lindner-2006-ChemPhysLett, Ashihara-2007-JPCA, Lindner-2007-ChemPhys, DeMarco-2016-JCP, Novelli-2020-JPCB, Novelli-2022-PCCP}.
It has been attributed to the lifetime of excited librations~\cite{Lindner-2006-ChemPhysLett, Lindner-2007-ChemPhys}
or to the development of the so-called ``hot ground state,'' where energy is transferred from high-frequency to low-frequency vibrational modes~\cite{DeMarco-2016-JCP}.
In stark contrast, other experimental and theoretical studies report librational relaxation as a much faster process, on the order of 0.2 ps~\cite{Ashihara-2007-JPCA},
supported by simulations~\cite{Yagasaki-2011-JCP-134, Petersen-2012-JPCB, Imoto-2015-JPCB}.
This discrepancy can be explained by differences in the simulation setup. 
In our simulations, the electric field acts globally, exciting all water molecules simultaneously.
Additionally, the applied pump pulse has a relatively long duration, continuously exciting librational modes over several pico seconds (see \fref{fig:pulse_and_temp}a). 
As a result, these modes are not only collectively excited but remain excited due to the sustained influx of energy.
Notably, previous simulations have also excited single molecules only and/or employed much shorter Gaussian pump widths, such as 47~fs~\cite{Yagasaki-2011-JCP-134, Petersen-2012-JPCB, Imoto-2015-JPCB}.
There, rotational degrees of freedom~-- particularly those of neighboring molecules~-- were identified to serve as the primary acceptors of energy, accounting for approximately 85~\% of energy redistribution~\cite{Petersen-2012-JPCB}.
In our setup, however, this rapid energy dissipation pathway is effectively blocked because all rotational degrees of freedom are already excited. 

Further evidence supporting widespread excitation of librational modes in our simulations comes from three observations:
(1) 
the birefringence $\Delta n$ shown in \fref{fig:pulse_and_temp}~(c) is quantitatively much larger than in experiments,
(2)
the maximum of the absorption coefficient is shifted to earlier pump–probe delay times, 
and  
(3)
the energy contained in rotational motion (see \fref{fig:temperatures}~(b) undergoes a maximum.
Although birefringence has been qualitatively observed in experiments using similar THz pulses~\cite{Novelli-2020-JPCB, Novelli-2022-PCCP}, 
the reported magnitude was orders of magnitude smaller. 
This can be attributed to the fact that in experiments, only a fraction of the water molecules is excited, whereas in our simulation, all molecules are affected. 
Since birefringence correlates with the population of anisotropically oriented dipole moments, this leads to an overestimated $\Delta n$ in our setup.
Finally, assuming first-order kinetics for the excitation of librational modes~-- such that the ground-state population depletes over the course of the excitation~-- also explains the observed shift of the absorption maximum to earlier delay times. 
As the available ground-state population diminishes, the excitation rate slows down even before the pump pulse reaches its peak.
This discussion underscores the importance of accurately modeling energy uptake and accounting for the duration and spatial characteristics of the pump pulse when making quantitative comparisons with experimental data. 
Nevertheless, it is reassuring that all qualitative experimental features, such as birefringence, are reproduced by our simulations. 

Having effectively disabled the relaxation pathway through neighboring rotational motion,
only the slower translational modes and the faster intramolecular vibrational modes
remain available for dissipating excess energy.
From \fref{fig:temperatures}~(b) and (c) we extract kinetic energy growth rates of \num{1.08}~ps and \num{1.96}~ps for translational and intramolecular motion, respectively.
These two channels, therefore, represent energy dissipation pathways with an effective combined time constant of approximately \num{0.70}~ps~-- strikingly similar to the relaxation time observed for the librational band. 
This suggests that the \num{0.70}~ps timescale corresponds to the dissipation of energy into all remaining modes, partially consistent with the hot ground state mechanism discussed above.
In contrast to its original interpretation, our simulations show that energy is not solely dissipated into slower translational motion, but also into faster vibrational motion, effectively heating all modes concurrently.
It is important to note, though, that equipartition does not necessarily hold for the individual components of atomic velocity (as also evident in \fref{fig:temperatures}b and c);
thus, a uniform increase in kinetic energy does not imply that the system has reached thermodynamic equilibrium.
Indeed, the observed temperature rise~-- on the order of $\approx 10^{13}$~K/s~-- appears unphysically fast. 
Remarkably, similar heating rates on the same order of magnitude have been inferred experimentally~\cite{DeMarco-2016-JCP}, 
giving rise to the notion of a ``hot ground state'': 
a long-lived non-equilibrium condition that relaxes toward equilibrium over much longer timescales. 
The temperatures shown in \fref{fig:temperatures} reflect the kinetic energy of the system, but do not imply equilibration on the potential energy surface, which is required for full thermal equilibrium.
It could thus very well be that our simulations have not reached equilibrium within the simulated time window.
This interpretation is further supported by \fref{fig:dynamics}, where the reorientational relaxation time in panel (a) has not yet converged to its equilibrium value.
While our data offers valuable insight into the formation of the hot ground state, we refrain from drawing definitive conclusions about its nature or dynamics. 
This is due to the fact that the relevant relaxation times extend beyond the accessible simulation window, and because the widespread excitation of librational modes, as discussed above, may fundamentally alter the relaxation behavior.%

It is noteworthy that only the relaxation time extracted from $\da$ (\num{0.79}~ps; see \fref{fig:pulse_and_temp}b) coincides with the time constant corresponding to the energy dissipation into translational and vibrational motion (\num{0.70}~ps; see \fref{fig:temperatures}b and c).
In contrast, 
the relaxation time extracted from $\dan$ is significantly shorter (\num{0.25}~ps).
This discrepancy suggests that the frequency-dependent refractive index $n(\tau_2, \omega)$ contributes substantially to the modulation of the spectrum~-- a conclusion also supported by the pronounced birefringence observed in \fref{fig:pulse_and_temp}c.
Technically this is an important observation: 
the simulations directly yield the product $\alpha(\tau_2, \omega)n(\tau_2, \omega)$ (see \eref{eq:absorption}), while $\alpha(\tau_2, \omega)$ must be separated off through postprocessing using the Kramers-Kronig relations~\cite{Iftimie-JCP-2005, Sun-2014-JACS}.
Given the magnitude of the birefringence observed herein (see \fref{fig:pulse_and_temp}c), it is conceivable that $\da$ and $\dan$ could become more similar when fewer molecules are excited (as discussed previously).
Nonetheless, the present result is a strong indication that relaxation times extracted from $\dan$ may be biased, and should be interpreted with caution~-- particularly when comparing to experimental data, where $\alpha(\omega)$ is the directly measured quantity.

A final point of interest is the system’s response to the applied THz pulse, which is relatively long, with a full width at half maximum (FWHM) of approximately \num{3}~ps.
We observe a distinct non-linear response in the kinetic energy associated with rotational motion (\fref{fig:temperatures}b), as well as in the H-bond lifetime and reorientational relaxation time shown in \fref{fig:dynamics}. 
Notably, the birefringence depicted in \fref{fig:pulse_and_temp} already indicates a non-linear optical response, consistent with observations from optical Kerr effect spectroscopy in liquid water~\cite{Elgabarty-2020-SciAdv}.
This transient anisotropy is further supported by the behavior of the H-bond network \fref{fig:dynamics}, indicating that the THz pulse temporarily perturbs the hydrogen-bond structure. 
These findings demonstrate that the effect of the THz pulse extends beyond simple heating and is sensitive to the specific shape and characteristics of the pulse~-- enabling pulse design and sequence engineering~\cite{Elgabarty-2020-SciAdv, Novelli-2020-JPCB, Novelli-2022-PCCP}.
Looking forward, it will be particularly interesting to explore how such extended THz pulses influence solute–solvent interactions and whether they could enable control over solvation behavior or actively modulate hydration dynamics.

\section{Conclusions}

In conclusion, building on our previous work on infrared spectroscopy and machine learning molecular dynamics (MLMD) under external electric fields, we have extended this framework to include time-dependent fields that mimic excitation pump pulses. 
This allows us to directly simulate the spectral response as a function of pump–probe delay time. 
Applying this methodology  to liquid water excited by a 12.3~THz pulse with a 3~ps width, we computed the time-resolved response by averaging over 1600 MLMD trajectories (20~ps each), totaling 32~ns of simulation data at \emph{ab initio} accuracy.
Our approach reproduces key experimental observables, including transient birefringence, relaxation times, energy redistribution pathways, and perturbations of the hydrogen-bond network. 
These results demonstrate that our method captures essential non-equilibrium dynamics in excellent agreement with experiment, where available.
Importantly, the method is not restricted to Gaussian excitation pulses or vibrational THz modes; 
it can be readily extended to arbitrary pulse shapes, multiple pulses, or other excitation frequencies. 
As such, the presented non-equilibrium MLMD approach opens the door to systematically investigating how external fields influence solvent dynamics and hydration behavior. 
This capability may prove especially valuable in exploring field-driven effects in reactive systems~-- impossible with force field MD due to fixed bonds and prohibitively expensive for full \emph{ab initio} MD.

\vspace{1cm}
\noindent\textbf{Acknowledgement}

We thank Fabio Novelli (Bochum, Southampton) for fruitful discussions.
This project was supported by the \textit{Deutsche Forschungsgemeinschaft} (DFG, German Research Foundation) under Germany's Excellence Strategy~-- EXC 2033~-- 390677874~-- RESOLV.

\clearpage
\printbibliography

\clearpage


\section*{Supplementary material}

# Contents

# S1 Evaluating Two-Point Correlation Functions

In equilibrium MD simulations, a discrete time correlation function is typically calculated as:

$$\langle A(0)A(t)\rangle = \frac{1}{N-t}\sum_{t_0=1}^{N-t} A_{t_0}\cdot A_{t_0+t}, \tag{S1}$$

where $A(t)$ is any time-dependent observable and $N$ the total number of frames in the trajectory. Here, $t_0$ spans all possible initial times such that $t_0 + t < N$, effectively averaging over all initial conditions in the trajectory. As a result, the correlation function depends only on the time difference $t$, not on the absolute initial time $t_0$. In non-equilibrium simulations, however, this equivalence of initial times no longer holds, and the correlation function becomes explicitly dependent on the initial condition $t_0$. Consequently, the ensemble average:

$$\langle A(t_0)A(t_0+t)\rangle = A_{t_0}\cdot A_{t_0+t}, \tag{S2}$$

collapses to a single term – clearly insufficient to represent the full ensemble. This challenge can be addressed by introducing a large number of independent trajectories and averaging over them. Following this approach, we generated 1600 trajectories in total. However, even with this ensemble size, directly averaging the two-point correlation function across trajectories provides insufficient statistical quality to yield well-converged spectra. To improve sampling, we introduce an additional averaging over a short time span $\Delta t$:

$$\langle A(t_0)A(t_0+t)\rangle = \frac{1}{\Delta t}\sum_{t_i=t_0-\Delta t/2}^{t_0+\Delta t/2} A_{t_i}\cdot A_{t_i+t}, \tag{S3}$$

assuming that $A(t)$ does not vary significantly within this window. In this sense, $\Delta t$ defines the temporal precision with which the *initial time* $t_0$ resolved, but leaves the temporal precision of $t$ unchanged. A smaller $\Delta t$ offers higher temporal resolution but requires a larger number of trajectories to maintain statistical accuracy. Notably, it is sensible to use different $\Delta t$ depending on the observable. All IR spectra presented are calculated with $\Delta t = 400$ fs, which provides a good compromise between statistical quality and temporal resolution. In case of the reorganization relaxation times and H-bond lifetimes we employed a smaller window of $\Delta t = 50$ fs. This is feasible because the statistics are improved by averaging over all water molecules and all H-bond pairs in the system, respectively. In principle, one could further reduce $\Delta t$ by increasing the number of trajectories. However, achieving a resolution of $\Delta t = 1$ frame would require many more than an order of magnitude increase in trajectory count. For the present study – covering 32 ns in total (1600 trajectories of 20 ps each) – this was deemed unnecessary, though such refinement may be required in future work.

## S2 Atomistic Decomposition of the Absorption Coefficient

The critical property required for the IR absorption coefficient (see Eq. 2 in the main text) is the time correlation function of the total dipole moment time derivative,

$$\left\langle \dot{\mathbf{M}}(\tau_2)\dot{\mathbf{M}}(\tau_2+t)\right\rangle. \tag{S4}$$

Therein, the total dipole moment time derivative

$$\dot{\mathbf{M}} = \sum_i \mathbf{P}_i(t)\mathbf{v}_i(t) \tag{S5}$$

can be replaced by the sum over all atoms $i$, multiplying the atomic polar tensor $\mathbf{P}_i(t)$ with the velocity $\mathbf{v}_i(t)$ of that atom. Inserting this relation into Eq. S4 yields Eq. 3 shown in the main text.

According to Eq. 8 in the main text, the velocity of each atom $i$

$$\mathbf{v}_i(t) = \sum_\zeta^{t,r,v} \mathbf{v}_{i,\zeta}(t) \tag{S6}$$

can further be decomposed into translational (t), rotational (r), and vibrational (v) components. Note that Eq. S6 is a short-hand notation of Eq. 8 in the main text. Inserting this sum into Eq. 3 yields

$$\left\langle \dot{\mathbf{M}}(\tau_2)\dot{\mathbf{M}}(\tau_2+t)\right\rangle = \left\langle \sum_i \mathbf{P}_i(\tau_2) \sum_\zeta^{t,r,v} \mathbf{v}_{i,\zeta}(\tau_2) \cdot \sum_j \mathbf{P}_j(\tau_2+t) \sum_\xi^{t,r,v} \mathbf{v}_{i,\xi}(\tau_2+t)\right\rangle \tag{S7}$$

where we now have a product of four sums, allowing us to rearrange the equation such that

$$\left\langle \dot{\mathbf{M}}(\tau_2)\dot{\mathbf{M}}(\tau_2+t)\right\rangle = \sum_{\zeta\xi}^{t,r,v} \left\langle \sum_i \mathbf{P}_i(\tau_2)\mathbf{v}_{i,\zeta}(\tau_2) \cdot \sum_j \mathbf{P}_j(\tau_2+t)\mathbf{v}_{i,\xi}(\tau_2+t)\right\rangle \tag{S8}$$

$$\equiv \sum_{\zeta\xi}^{t,r,v} C_{\zeta\xi}(\tau_2,t) \tag{S9}$$

the sum over translations, rotations, and vibrations is now outside of the time correlation function and we introduced the short-hand notation $C_{\zeta\xi}(\tau_2,t)$. Since calculating a time correlation function is a linear operation, see Eq. S1, that operation is allowed. Note that the sums running over all atoms $i$ and $j$ remain in the time correlation function, i.e. are being averaged over, but it is also possible to obtain the spectral contribution for each individual atom as well, which is reminiscent of the cross correlation analysis [1]. Finally, inserting Eq. S9 into the basic equation for the absorption coefficient (Eq. 2 in the main text) yields

$$\alpha(\tau_2,\omega) = \frac{\beta\pi}{3cV\epsilon_0 n(\tau_2,\omega)} \int_{-\infty}^{\infty} e^{-i\omega t} \sum_{\zeta\xi}^{t,r,v} C_{\zeta\xi}(\tau_2,t)\,dt. \tag{S10}$$

Since the Fourier transform is again a linear operation, the sum can again be taken out of the integral

$$\alpha(\tau_2,\omega) = \sum_{\zeta\xi}^{t,r,v} \frac{\beta\pi}{3cV\epsilon_0 n(\tau_2,\omega)} \int_{-\infty}^{\infty} e^{-i\omega t} C_{\zeta\xi}(\tau_2,t)\,dt \equiv \sum_{\zeta\xi}^{t,r,v} \alpha_{\zeta\xi}(\tau_2,\omega) \tag{S11}$$

recovering Eq. 9 in the main text. Note that the frequency-dependent refractive index $n(\tau_2,\omega)$ is not decomposed, but computed once from the total spectrum $\alpha(\tau_2,\omega)$ using the Kramers-Kronig relations [2, 3]. By dividing by $n(\tau_2,\omega)$ each individual $\alpha_{\zeta\xi}(\tau_2,\omega)$ can be obtained.

## S3 IR Spectra at pump-probe delays of $-5$ and $5$ ps

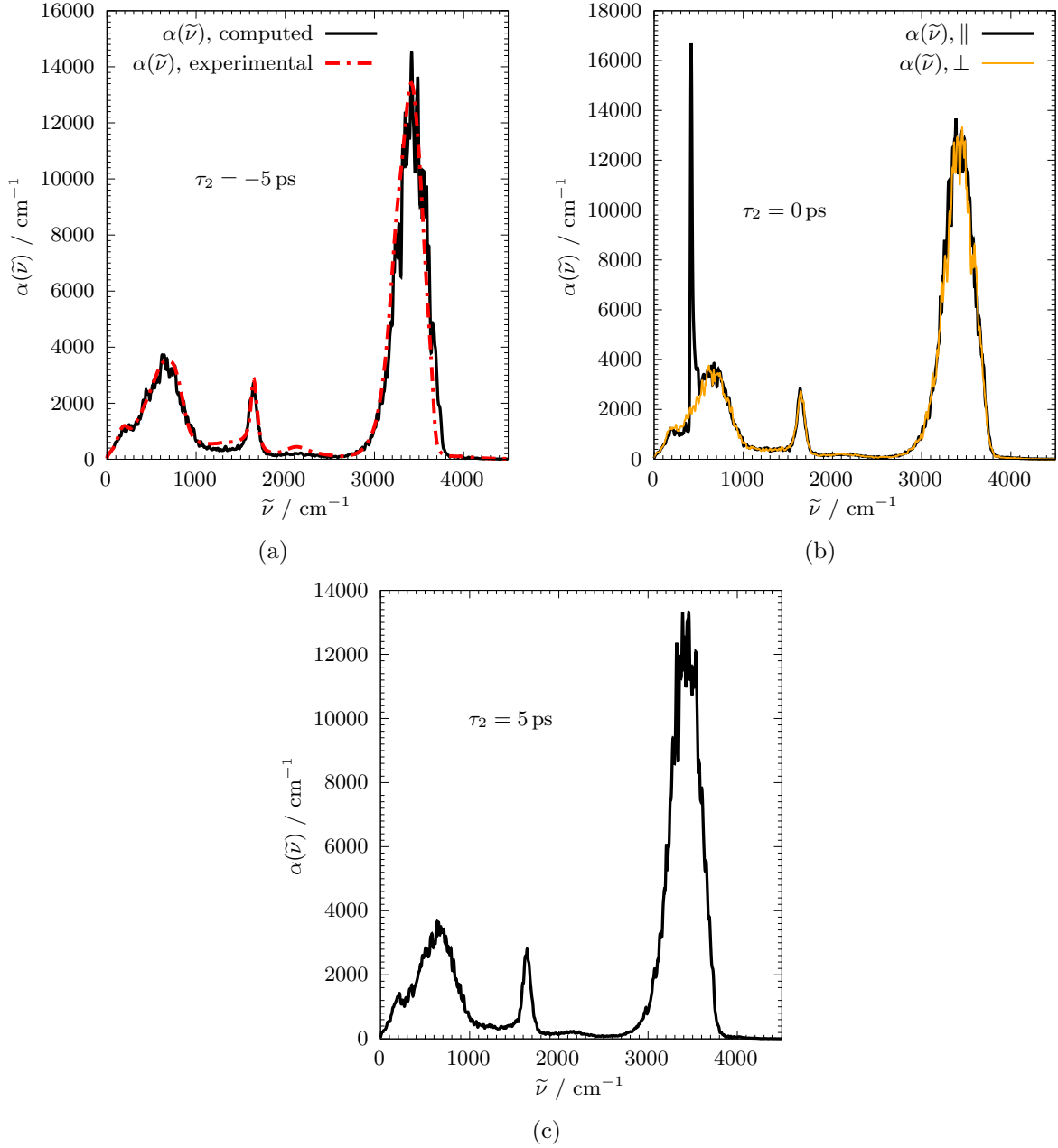

(a)

(b)

(c)

Figure S1: IR absorption coefficients parallel to the electric field of the pump pulse (black solid lines) calculated from the simulations presented herein using Eq. 2 in the main text at a pump–probe delay $\tau_2$ of $-5$ (a), 0 (b), and 5 ps (c), corresponding to times before, during, and after the pump pulse, respectively. The delay time of 0 ps (b) coincides with the maximum of the pump pulse. To illustrate the resulting transient dichroism induced by the pump, the absorption coefficient perpendicular to the electric field of the pump is also shown (orange solid line). The two spectra before (a) and after the pulse (c) are indistinguishable within statistical uncertainty, despite a total of 32 ns of trajectory data being used. The red dashed–dotted line in panel (a) shows experimental reference data from Ref. [4], highlighting the accuracy of the present method in reproducing the IR spectrum of liquid water at ambient conditions across the full range from 10 to $4500\,\mathrm{cm}^{-1}$ – including the excitation frequency of 12.3 THz ($\approx 410\,\mathrm{cm}^{-1}$). Note that no artificial frequency shifts or intensity scaling have been applied.

## S4 Frequency-Dependent Refractive Index

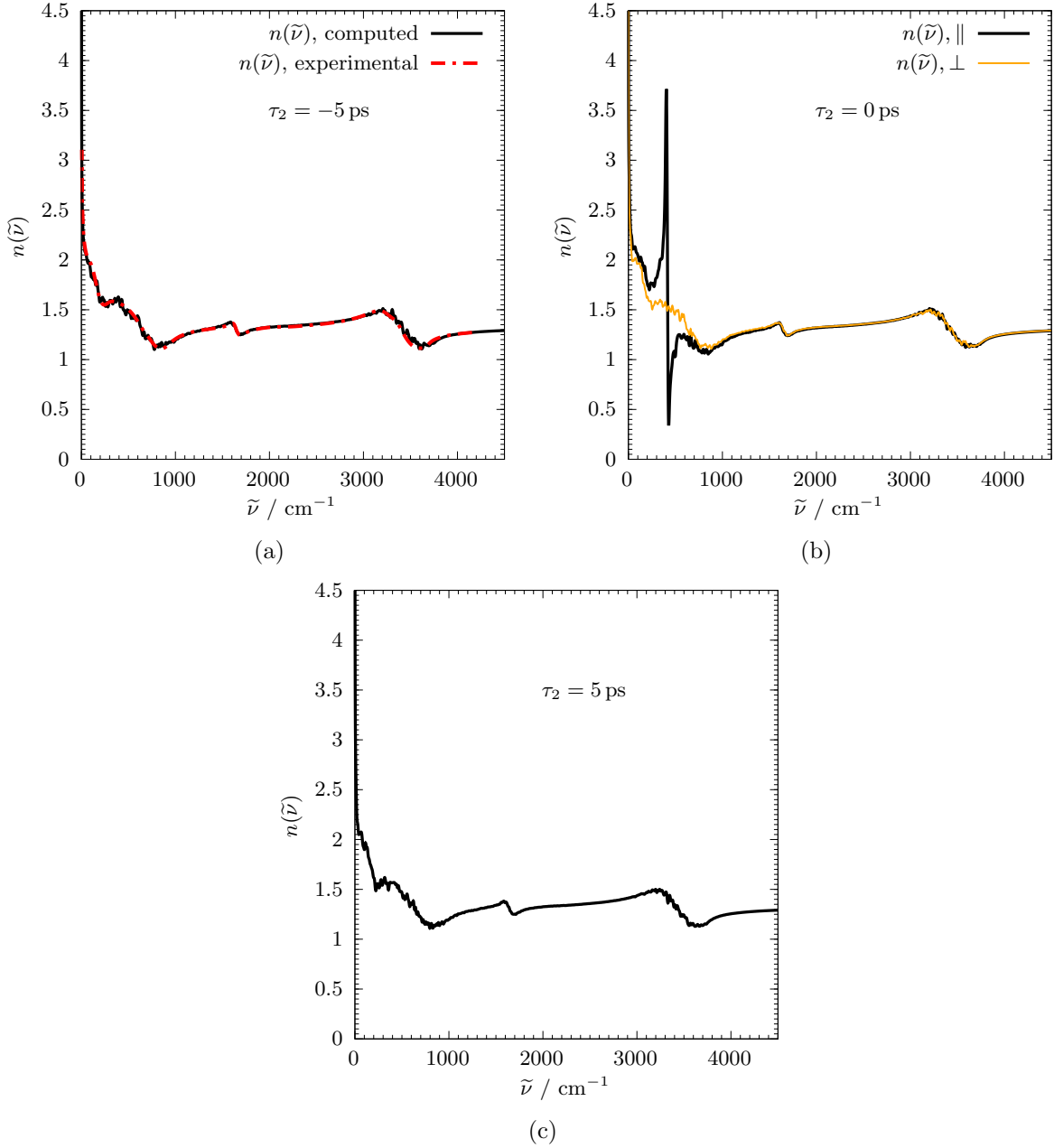

Figure S2: Frequency-dependent refractive indices parallel to the electric field of the pump pulse (black solid lines) calculated from the simulations presented herein at a pump-probe delay $\tau_2$ of $-5$ (a), 0 (b), and 5 ps (c), corresponding to times before, during, and after the pump pulse, respectively. The delay time of 0 ps (b) coincides with the maximum of the pump pulse. To illustrate the resulting transient birefringence induced by the pump, the frequency dependent refractive index perpendicular to the electric field of the pump is also shown (orange solid line). The two frequency-dependent refractive indices before (a) and after the pulse (c) are indistinguishable within statistical uncertainty, despite a total of 32 ns of trajectory data being used. The red dashed–dotted line in panel (a) shows experimental reference data from Ref. [4], highlighting the accuracy of the present method in reproducing the frequency dependent refractive index of liquid water at ambient conditions across the full range from 10 to $4500\,\mathrm{cm}^{-1}$ – including the excitation frequency of 12.3 THz ($\approx 410\,\mathrm{cm}^{-1}$).



\end{document}